\documentclass[prl,showpacs,preprintnumbers,amsmath,amssymb,endfloats]{revtex4}

\usepackage{graphicx}
\usepackage{dcolumn}
\usepackage{bm}
\usepackage{epstopdf}
\usepackage{latexsym}
  
\newcommand{\be}{\begin{equation}}
\newcommand{\ee}{\end{equation}}
\newcommand{\bea}{\begin{eqnarray}}
\newcommand{\eea}{\end{eqnarray}}

\newcommand{\req}[1]{Eq.~(\ref{#1})}

\begin{document}


\title{Interface Spin-Orbit Coupling in a Non-centrosymmetric Thin-Film Superconductor} 

\author{X. S. Wu and P. W. Adams}
\affiliation{Department of Physics and Astronomy\\Louisiana State University\\Baton Rouge, Louisiana, 70803}%

\author{Y. Yang and R. L. McCarley}
\affiliation{Department of Chemistry and the Center for Biomodular Multi-scale Systems\\Louisiana State
University\\Baton Rouge, LA, 70803}

\date{\today}

\begin{abstract}
We present a detailed study of the effects of interface spin-orbit coupling (ISOC) on the critical field behavior of non-centrosymmetric (NCS), ultra-thin superconducting Be/Au bilayers.  Parallel field measurements were made in bilayers with Be thicknesses in the range of $d=2 - 30$ nm and Au coverages of 0.5 nm.  Though the Au had no significant effect on the superconducting gap, it produced profound changes in the spin states of the system.  In particular, the parallel critical field exceeded the Clogston limit by an order of magnitude in the thinnest films studied.  In addition, the parallel critical field unexpectedly scaled as $H_{c||}/\Delta_o\propto1/d$ suggesting that the spin-orbit coupling energy was proportional to  $\Delta_o/d^2$.   Tilted field measurements showed that contrary to recent theory, the ISOC induces a large in-plane superconducting susceptibility but only a very small transverse susceptibility.  
\end{abstract}

\pacs{74.20.Rp,74.78.Db,73.40.Jn}
\maketitle
 
One of the most fundamental characteristics of a superconductor is the symmetry of its condensate wavefunction.  Indeed, this has proven to be a central issue in the description of a number of non-conventional superconductors such as high-T$_c$ systems \cite{Tsuei2000} and the ruthenates \cite{Mackenzie2003,Liu2004}.  In conventional BCS superconductivity the condensate is time reversal invariant and is formed from Cooper pairs consisting of electrons of opposite momentum and opposite spin \cite{Tinkham1980}.  In non-conventional superconductors, however, this simple symmetry can be modified by the underlying crystal structure and/or the symmetry of the pairing interaction.  A compelling example of the former is the recently discovered heavy fermion superconductor CePt$_3$Si, whose crystal structure lacks inversion symmetry \cite{Bauer2004}.  CePt$_3$Si exhibits a line node gap structure \cite{Onuki2005,Bauer2005,Young2005} which is believed to be, in part, a consequence of strong spin-orbit (SO) coupling in a non-centrosymmetric (NCS) crystal symmetry \cite{Bauer2004,Frigeri2004a,Samokhin2004a}.   This has stimulated renewed interest in the possibility of realizing non-conventional pairing states from the convolution of broken inversion symmetry and spin orbit coupling, neither of which violate time reversal invariance.  Recent theoretical work on the problem has suggested that a bilayer configuration, which, by definition, lacks inversion symmetry, consisting of a low atomic mass thin-film superconductor coated with a high atomic mass non-superconducting element such as Au or Pt would be a model system for studying the effects of SO coupling in a NCS superconductor \cite{Edelstein2003,Gor'kov2001,Yip2002}.   In the present Letter, we present a study of interface SO coupling in thin Be films coated with 0.5 nm of Au via critical magnetic field measurements.  Contrary to recent theoretical predictions, we find that the SO induced in-plane superconducting spin susceptibility is significantly larger than the corresponding perpendicular susceptibility.  Furthermore, the SO coupling cannot be described in terms of Abrikosov-Gorkov impurity formalism \cite{Tinkham1980} in that it appears to be a function of the superconducting gap.

In the experiments described below we use critical field measurements to determine the SO coupling strength in Be/Au films of varying Be thickness.  The Maki equation \cite{Maki1966,AOI1974} is a useful tool for extracting the spin response of the superconductor from the orbital response, particularly when the film is not in the thin-film limit and/or the field is not parallel to the film surface.  In general the critical field of a thin film is a function of the superconducting gap $\Delta_o$, the film thickness $d$, the electron diffusivity $D$,  and the spin orbit coupling parameter $b$.  The critical field $H_c$ is determined by the implicit function \cite{AOI1974}:
\begin{equation}
\ln\left(\frac{T}{T_c}\right)=\psi\left(\frac{1}{2}\right)-\frac{\alpha_+}{2\gamma}\psi\left(\frac{1}{2}+\frac{\epsilon +2\alpha_-}{4\pi k_BT}\right)
+\frac{\alpha_-}{2\gamma}\psi\left(\frac{1}{2}+\frac{\epsilon +2\alpha_+}{4\pi k_BT}\right)
\label{Maki}
\end{equation}
where
\begin{equation}
\alpha_{\pm}=b\pm \gamma, \gamma=(b^2-\mu_B^2H_c^2)^{1/2}\nonumber, 
\label{alpha}
\end{equation}
$T_c$ is the critical temperature, and $\psi$ is the digamma function. $\epsilon$ is a function of the angle between the plane of the film and the magnetic field,
\begin{equation}
\epsilon(\theta)=D[2eH_c\sin(\theta)+\frac{1}{3}(deH_c\cos(\theta))^2/\hbar]\nonumber
\label{epsilon}
\end{equation}
where $\theta=0$ corresponds to field parallel to the film plane. The parallel field solutions of \req{Maki} in the $d\rightarrow0$ limit, where orbital effects are negligible, are of particular relevance \cite{Tinkham1980},
\bea
\frac{2\mu_BH_{c||}}{\Delta_o}\approx&\sqrt{2}&\;\;\;b/\Delta_o\ll1 \label{eq:Clogston} \\
=&\sqrt{3b/\Delta_o}&\;\;\;b/\Delta_o\geq1,
\label{eq:Zeeman}
\eea
where \req{eq:Clogston} is the familiar Clogston spin-paramagnetic critical field \cite{Clogston1962,Chandrasekhar1962,SPsolution}.  Note that from \req{eq:Zeeman} the critical field can be arbitrarily high with increasing $b$.  In contrast, if the Zeeman coupling is neglected then, at any finite thickness $d<\xi$, the parallel critical field is limited by the orbital term,
\be
\frac{2\mu_BH_{c||}}{\Delta_o}=\sqrt{\frac{3\hbar^3}{m^2D\Delta_o}}\frac{1}{d}
\label{orbital}
\ee

Numerous studies of the spin-paramagnetic transition in ultra-thin Al and Be films have shown that these two light elements have a very low intrinsic SO scattering rate \cite{Tedrow1979,Adams2004,Adams1998} and are true spin-singlet superconductors.  Consequently, they make ideal candidates for systematic studies of the effects of ISOC induced by impurity coatings \cite{Tedrow1982,Edelstein2003}.  Not only does this bilayer configuration break inversion symmetry, but one can treat the ISOC as a boundary condition on the spin component of the superconducting wave function.  Early measurements of the critical field behavior of Al films coated with less than 0.2 nm of Pt revealed that the Pt-induced ISOC can produce significant enhancements of the parallel critical field with virtually no effect on the transition temperature \cite{Tedrow1979}.   Though the temperature and angular dependencies of the Al/Pt critical fields were in reasonable agreement with the Maki formula, the overall magnitude of $H_{c||}$ was typically $\sim10-30\%$ higher than could be accounted for by reasonable estimates of $b$.  

Recent analysis of the spin states of two-dimensional NCS superconductors, and, in particular, superconducting-normal metal bilayers, predicts that ISOC will introduce an anisotropic spin triplet component into superconducting ground state \cite{Gor'kov2001,Yip2002}.  The triplet component is most directly manifest in the spin susceptibility of the superconductor $\chi^s$.  Of course, conventional spin-singlet BSC superconductors have $\chi^s\sim~0$ at $T=0$, but ISOC is expected to lift the spin degeneracy, and in the strong coupling l: $\chi^{s}_{\parallel}\sim\chi^n/2$, $\chi^{s}_{\bot}\sim\chi^n$, where $\chi^{s}_{\parallel}$ is the in-plane superconducting susceptibility and $\chi^n$ is the normal state susceptibility.  The Clogston critical field given by \req{eq:Clogston} is derived by equating the magnetic free energy {\it difference} between normal and superconducting states with the condensation energy of the superconducting state.  If $\chi^s\neq0$ then \req{eq:Clogston} can be generalized,
\be
2\mu_BH_c=\sqrt{2}\Delta_o/\sqrt{1-\chi^s/\chi^n}.
\label{GenClogston}
\ee
This would imply that the parallel critical field of a bilayer, such as the Al/Pt samples of Ref.\ \onlinecite{Tedrow1979}, would never be larger than $H_{c||}=\Delta_o/\mu_B$.   However, the critical field enhancements observed in those early experiments were significantly greater than this upper limit.  Since the late 1970's  very little work has been done on SO coupling in thin film superconductors, though several fundamental questions remain unanswered.  For instance, the Al/Pt measurements were made at a single Al thickness and over a very narrow range of gap values.   Clearly, one would like to determine the depth to which the ISOC penetrates into the superconducting film and how the spin component of critical field behavior scales with film thickness and gap magnitude.  Another important issue is whether or not the anisotropy in $\chi^s$ is observable in the angular dependence of the critical field. 

Be/Au bilayer films of varying Be thickness were prepared by e-beam evaporation in a initial vacuum of $\sim0.2\mu$ Torr. All of the depositions were made on fire polished glass substrates held at 84 K.  First a layer of Be film with thickness in the range $2. - 30.0$ nm was deposited at a rate of 0.14 nm/s, then a 0.5 nm Au film was deposited at 0.01 nm/s without breaking the vacuum.  Both the Be and Be/Au films were found to be very smooth and homogenous,  with no evidence of islanding or granularity, see Fig. \ref{AFM}. The films were subsequently trimmed in order to eliminate edge effects.  Resistive measurements were made in a dilution refrigerator with a base temperature 50 mK by a standard four-probe lock-in method.  The films were aligned with the magnetic field via an {\it in situ} mechanical rotator.  In the data presented below the transition temperatures were defined by the temperature at which the resistance fell to 10\% of its normal state value and the critical field was determined by the midpoint of the resistive transition.
 
The transition temperature of the homogeneously disordered Be/Au films used in this study are plotted as a function of Be thickness in Fig.\ \ref{Tc&D}.  Films with $d<2$ nm are known to display a non-perturbative zero bias anomaly in their tunneling density of states, which is associated with the emergence of the Coulomb gap \cite{Adams2000b}.  As can be seen in Fig.\ \ref{Tc&D}, this is also the critical thickness below which the zero temperature superconducting phase is lost and the electron diffusivity goes to zero.  In order to make use of \req{Maki} it was necessary to fit the thickness dependence of $T_c$ and $D$ with an empirical functional form.  In particular, the solid lines in the main panel and the inset of Fig.\ \ref{Tc&D} are the functions: 
\begin{eqnarray}
T_c(d)&=&T_{co}\tanh[(d-1.35)/1.29)] \label{empirical:Tc}\\
D(d)&=&D_o\tanh[d^2/23.3],
\label{empirical:D}
\eea
where $T_{co}=0.68$ K and $D_o\sim3\hbar/m$.  $D_o$ was determined from films with $d>10$ nm using the relation $1/R=2e^2\nu_0D_od$, where $R$ is the sheet resistance and $\nu_0$ is the density of states per spin of Be.  The BCS coherence length for a Be film with $T_c\sim0.7$ K is $\xi_o\sim4$ $\mu m$.  For the range of diffusivities plotted in the inset of Fig.\ \ref{Tc&D} the mean free path is always $l_o<1$ nm and the corresponding Pippard coherence lengths are in the range $\xi=0.85\sqrt{\xi_ol_o}\approx 20 - 40$ nm.  Consequently, all of the data discussed below is in the  `dirty' limit where $l_o\ll d<\xi$. 

We have measured the parallel critical field $H_{c||}$ of Be films of varying thickness ($d\approx2 - 30$ nm) with and without 0.5 nm Au overlayers.  The Au coatings did not significantly affect $T_c$ but did $\it increase$ the normal state resistances by $10-50\%$.  Since the transition temperature of the films varied greatly over this thickness range, we normalized the critical fields by the superconducting gap.  In Fig.\ \ref{Linear} we plot $2\mu_B H_{c||}/\Delta_o$ as a function of the inverse Be thickness at 60 mK, where we used the relation $\Delta_o=2.1k_BT_c$, as determined by tunneling density of states measurements \cite{Adams2004}.  The triangular symbols correspond to pure Be films.  Note that, except for the thickest Be sample, the critical field is independent of $d$ and precisely that of \req{eq:Clogston}, indicating that orbital effects were negligible in films of thickness less than $\sim10$ nm.  In contrast, the normalized critical field of the Be/Au bilayers (circular symbols) is not only a strong function of thickness, but it exceeds the Clogston limit by more than a factor of 8 in the thinnest film!  Furthermore, as can be seen by the solid line in Fig.\ \ref{Linear}, $2\mu_B H_{c||}/\Delta_o$ is linear in $1/d$ across the entire range of thicknesses.   The dashed line in Fig.\ \ref{Linear} represents the thickness dependence of the orbital critical field of \req{orbital} using Eqs.\ (\ref{empirical:Tc}) and (\ref{empirical:D}).  Clearly, the critical field of bilayers with $d<5$ nm is Zeeman mediated giving us a direct probe of the ISOC.  

In the context of \req{Maki} one would attribute the linear critical field behavior in Fig.\ \ref{Linear} to the thickness dependence coupling parameter $b$.   In a related study Bergmann \cite{BERGMANN1985,Bergmann2001} used magnetoresistance measurements to determine the SO scattering rate in non-superconducting Mg films coated with submonolayer coverages of Au.  The Au produced large increases in an otherwise small intrinsic Mg SO scattering rate.  In fact, the SO rate was observed to scale as $b\sim 1/d$.  This scaling behavior, however, is incompatible with the linear behavior in Fig.\ \ref{Linear}.   If one assumes that \req{eq:Zeeman} is applicable for the thinnest bilayers, $d\le5$ nm, then the SO coupling parameter must scale as $b\sim\Delta_o/d^2$ in order for the normalized critical field to be proportional to $1/d$.  This suggests that the ISOC is not simply a scattering process.  In fact, ISOC does not break time reversal symmetry, and is therefore not a pair-breaking interaction.  Consequently, it is not meaningful to cast ISOC in terms os scattering processes, particularly in the $T,H\rightarrow0$ limit.   Alternatively, it seems more appropriate to characterize ISOC as a boundary condition on the spin states of the superconductor.  In a semi-infinite slab, for instance, the superconductor would have a non-negligible spin-triplet component near the Be-Au interface but decay into a pure spin singlet eigenstate as one moves into the interior.   

As discussed above, the spin susceptibility of the NCS bilayer is expected to be anisotropic in the strong ISOC limit.  In order to address this issue we have measured the critical field of Be films and Be/Au bilayers as a function of tilt angle between the magnetic field and the plane of the film.   Upon tilting the sample out of parallel orientation, orbital contributions to the critical field quickly dominate.  Consequently, it is difficult to infer any anisotropy in the Zeeman response from the raw rotational data (see insets of Fig.\ \ref{DipFit}).  To circumvent this, we have measured the ratio of the critical field of Be/Au bilayers and Be films of equal thickness as a function of tilt angle $\theta$.  This, in effect, greatly improves the sensitivity to variations on the orbital background, thus giving us access to the angular dependence of the Zeeman contribution.  Typical behavior is shown in Fig.\ \ref{Dip}.  In order to better display the overall structure of the curves we have normalized the parallel field ratios to unity in this plot.  Note the dip structure near $20^\circ$ in each curve.  We believe that this structure is consequence of an anisotropic susceptibility.  In Fig.\ \ref{DipFit} we plot the ratio of the critical field of a 5.4 nm Be/Au bilayer and a 5.4 nm Be film as a function of angle.  The perpendicular critical field $H_{c2}$ of the bilayer was a factor of 3 higher than that of the Be film, hence the ratio is not unity at $\theta=90^\circ$.  The solid line in Fig.\ \ref{DipFit} is the expected angular dependence of \req{Maki}, using an isotropic SO coupling parameter of $b=0.013$ mV for the Be film and $b=1.85$ mV for the bilayer, as determined from fits to the inset data.  For both samples $\Delta_o\sim0.1 mV$.  An isotropic $b$ leads to a monotonic interpolation between the parallel and perpendicular field ratios.  The isotropic solution is, in fact, monotonic independent of the relative thickness, $T_c$, resistance, and/or perpendicular critical fields of the two films.  This reflects the fact that it is almost indistinguishable from what one would obtain using the phenomenological Tinkham formula \cite{Tinkham1980,AOI1974} which only involves $H_{c||}$ and $H_{c2}$.  In contrast, the dashed line depicts the solution to \req{Maki} assuming an exponentially attenuated SO parameter $b=1.85\exp(-\theta/\theta_o)$ mV, with a characteristic angle $\theta_o=2.5^\circ$.  The dashed line captures the dip structure in the data which indicates that the Zeeman component of the critical field is angle dependent.  The dip structure is consistent with an anisotropic superconducting susceptibility consisting of a large in-plane component (of the order of the normal state susceptibility) and a small transverse component.  The sense of this anisotropy is of the opposite sign of that calculated for non-disordered 2D NCS superconductors, where the in-plane component is expected to be half that of the transverse component.  Nevertheless, the fact that ISOC induces a preferentially large in-plane susceptibility would also account for earlier reports of anomalously high parallel critical fields in Al/Pt bilayers \cite{Tedrow1979}.  

In summary, we have used critical field measurements to probe the effects of interface spin-orbit coupling on the spin states of superconducting Be films.  We find that a $\sim$1 monolayer coating of Au on the Be surface produces a large, anisotropic enhancement to the Zeeman component of the critical field.  The normalized parallel critical field $H_{c||}/\Delta_o$ of Be/Au bilayers scales as the inverse of the Be thickness and can exceed the Clogston limit by as much as an order of magnitude.  The critical field scaling was observed over a wide range of $\Delta_o$ which is inconsistent with the standard impurity scattering model of SO coupling.  We believe that the scaling is, in fact, a manifestation of the superconductor's attempt to reconcile a mixed-spin boundary condition at the Au interface with the intrinsic spin-singlet ground state of Be.  Naively, one would expect that the ISOC healing length would be of the order of $\xi$, but it may be significantly shorter in the presence of disorder.  Nevertheless, the anisotropic spin susceptibility is clearly evident in the tilted field data.  A high field study of the local tunneling density of states at the Be/Au and the Be/substrate interfaces, respectively, should provide an important local probe of the extent of spin-mixing at the two boundaries.  

We gratefully acknowledge enlightening discussions with Victor Edelstein, Gianluigi Catelani, Ilya Vekhter, Dana Browne, and David Young. This work was supported by the National Science Foundation under Grant DMR 02-04871.

\bibliographystyle{apsrev}
\bibliography{SpinOrbit,Triplet,PairingResonance}

\begin{thebibliography}{25}
\expandafter\ifx\csname natexlab\endcsname\relax\def\natexlab#1{#1}\fi
\expandafter\ifx\csname bibnamefont\endcsname\relax
  \def\bibnamefont#1{#1}\fi
\expandafter\ifx\csname bibfnamefont\endcsname\relax
  \def\bibfnamefont#1{#1}\fi
\expandafter\ifx\csname citenamefont\endcsname\relax
  \def\citenamefont#1{#1}\fi
\expandafter\ifx\csname url\endcsname\relax
  \def\url#1{\texttt{#1}}\fi
\expandafter\ifx\csname urlprefix\endcsname\relax\def\urlprefix{URL }\fi
\providecommand{\bibinfo}[2]{#2}
\providecommand{\eprint}[2][]{\url{#2}}

\bibitem[{\citenamefont{Tsuei and Kirtley}(2000)}]{Tsuei2000}
\bibinfo{author}{\bibfnamefont{C.~C.} \bibnamefont{Tsuei}} \bibnamefont{and}
  \bibinfo{author}{\bibfnamefont{J.~R.} \bibnamefont{Kirtley}},
  \bibinfo{journal}{Rev. Mod. Phys.} \textbf{\bibinfo{volume}{72}},
  \bibinfo{pages}{969} (\bibinfo{year}{2000}).

\bibitem[{\citenamefont{Mackenzie and Maeno}(2003)}]{Mackenzie2003}
\bibinfo{author}{\bibfnamefont{A.~P.} \bibnamefont{Mackenzie}}
  \bibnamefont{and} \bibinfo{author}{\bibfnamefont{Y.}~\bibnamefont{Maeno}},
  \bibinfo{journal}{Rev. Mod. Phys.} \textbf{\bibinfo{volume}{75}},
  \bibinfo{pages}{657} (\bibinfo{year}{2003}).

\bibitem[{\citenamefont{Nelson et~al.}(2004)\citenamefont{Nelson, Mao, Maeno,
  and Liu}}]{Liu2004}
\bibinfo{author}{\bibfnamefont{K.~D.} \bibnamefont{Nelson}},
  \bibinfo{author}{\bibfnamefont{Z.~Q.} \bibnamefont{Mao}},
  \bibinfo{author}{\bibfnamefont{Y.}~\bibnamefont{Maeno}}, \bibnamefont{and}
  \bibinfo{author}{\bibfnamefont{Y.}~\bibnamefont{Liu}},
  \bibinfo{journal}{Science} \textbf{\bibinfo{volume}{306}},
  \bibinfo{pages}{1151} (\bibinfo{year}{2004}).

\bibitem[{\citenamefont{Tinkham}(1996)}]{Tinkham1980}
\bibinfo{author}{\bibfnamefont{M.}~\bibnamefont{Tinkham}},
  \emph{\bibinfo{title}{Introduction to Superconductivity}}
  (\bibinfo{publisher}{McGraw-Hill, New York}, \bibinfo{year}{1996}).

\bibitem[{\citenamefont{Bauer et~al.}(2004)\citenamefont{Bauer, Hilscher,
  Michor, Paul, Scheidt, Gribanov, Seropegin, Noel, Sigrist, and
  Rogl}}]{Bauer2004}
\bibinfo{author}{\bibfnamefont{E.}~\bibnamefont{Bauer}},
  \bibinfo{author}{\bibfnamefont{G.}~\bibnamefont{Hilscher}},
  \bibinfo{author}{\bibfnamefont{H.}~\bibnamefont{Michor}},
  \bibinfo{author}{\bibfnamefont{C.}~\bibnamefont{Paul}},
  \bibinfo{author}{\bibfnamefont{E.~W.} \bibnamefont{Scheidt}},
  \bibinfo{author}{\bibfnamefont{A.}~\bibnamefont{Gribanov}},
  \bibinfo{author}{\bibfnamefont{Y.}~\bibnamefont{Seropegin}},
  \bibinfo{author}{\bibfnamefont{H.}~\bibnamefont{Noel}},
  \bibinfo{author}{\bibfnamefont{M.}~\bibnamefont{Sigrist}}, \bibnamefont{and}
  \bibinfo{author}{\bibfnamefont{P.}~\bibnamefont{Rogl}},
  \bibinfo{journal}{Phys. Rev. Lett.} \textbf{\bibinfo{volume}{92}},
  \bibinfo{pages}{027003} (\bibinfo{year}{2004}).

\bibitem[{\citenamefont{Izawa et~al.}(2005)\citenamefont{Izawa, Kasahara,
  Matsuda, Yasuda, Settai, and Onuki}}]{Onuki2005}
\bibinfo{author}{\bibfnamefont{K.}~\bibnamefont{Izawa}},
  \bibinfo{author}{\bibfnamefont{Y.}~\bibnamefont{Kasahara}},
  \bibinfo{author}{\bibfnamefont{K.}~\bibnamefont{Matsuda},
  \bibfnamefont{Y.and~Behnia}},
  \bibinfo{author}{\bibfnamefont{T.}~\bibnamefont{Yasuda}},
  \bibinfo{author}{\bibfnamefont{R.}~\bibnamefont{Settai}}, \bibnamefont{and}
  \bibinfo{author}{\bibfnamefont{Y.}~\bibnamefont{Onuki}},
  \bibinfo{journal}{Phys. Rev. Lett.} \textbf{\bibinfo{volume}{94}},
  \bibinfo{pages}{197002} (\bibinfo{year}{2005}).

\bibitem[{\citenamefont{Bonalde et~al.}(2005)\citenamefont{Bonalde,
  Bramer-Escamilla, and Bauer}}]{Bauer2005}
\bibinfo{author}{\bibfnamefont{I.}~\bibnamefont{Bonalde}},
  \bibinfo{author}{\bibfnamefont{W.}~\bibnamefont{Bramer-Escamilla}},
  \bibnamefont{and} \bibinfo{author}{\bibfnamefont{E.}~\bibnamefont{Bauer}},
  \bibinfo{journal}{Phys. Rev. Lett.} \textbf{\bibinfo{volume}{94}},
  \bibinfo{pages}{207002} (\bibinfo{year}{2005}).

\bibitem[{\citenamefont{Young et~al.}(2005)\citenamefont{Young, Moldovan, Wu,
  Adams, and Chan}}]{Young2005}
\bibinfo{author}{\bibfnamefont{D.~P.} \bibnamefont{Young}},
  \bibinfo{author}{\bibfnamefont{M.}~\bibnamefont{Moldovan}},
  \bibinfo{author}{\bibfnamefont{X.~S.} \bibnamefont{Wu}},
  \bibinfo{author}{\bibfnamefont{P.~W.} \bibnamefont{Adams}}, \bibnamefont{and}
  \bibinfo{author}{\bibfnamefont{J.~Y.} \bibnamefont{Chan}},
  \bibinfo{journal}{Phys. Rev. Lett.} \textbf{\bibinfo{volume}{94}},
  \bibinfo{pages}{107001} (\bibinfo{year}{2005}).

\bibitem[{\citenamefont{Frigeri et~al.}(2004)\citenamefont{Frigeri, Agterberg,
  Koga, and Sigrist}}]{Frigeri2004a}
\bibinfo{author}{\bibfnamefont{P.~A.} \bibnamefont{Frigeri}},
  \bibinfo{author}{\bibfnamefont{D.~F.} \bibnamefont{Agterberg}},
  \bibinfo{author}{\bibfnamefont{A.}~\bibnamefont{Koga}}, \bibnamefont{and}
  \bibinfo{author}{\bibfnamefont{M.}~\bibnamefont{Sigrist}},
  \bibinfo{journal}{Phys. Rev. Lett.} \textbf{\bibinfo{volume}{92}},
  \bibinfo{pages}{097001} (\bibinfo{year}{2004}).

\bibitem[{\citenamefont{Samokhin et~al.}(2004)\citenamefont{Samokhin, Zijlstra,
  and Bose}}]{Samokhin2004a}
\bibinfo{author}{\bibfnamefont{K.}~\bibnamefont{Samokhin}},
  \bibinfo{author}{\bibfnamefont{E.}~\bibnamefont{Zijlstra}}, \bibnamefont{and}
  \bibinfo{author}{\bibfnamefont{S.}~\bibnamefont{Bose}},
  \bibinfo{journal}{Phys. Rev. B} \textbf{\bibinfo{volume}{69}},
  \bibinfo{pages}{094514} (\bibinfo{year}{2004}).

\bibitem[{\citenamefont{Edelstein}(2003)}]{Edelstein2003}
\bibinfo{author}{\bibfnamefont{V.~M.} \bibnamefont{Edelstein}},
  \bibinfo{journal}{Phys. Rev. B} \textbf{\bibinfo{volume}{67}},
  \bibinfo{pages}{020505} (\bibinfo{year}{2003}).

\bibitem[{\citenamefont{Gor'kov and Rashba}(2001)}]{Gor'kov2001}
\bibinfo{author}{\bibfnamefont{L.~P.} \bibnamefont{Gor'kov}} \bibnamefont{and}
  \bibinfo{author}{\bibfnamefont{E.~I.} \bibnamefont{Rashba}},
  \bibinfo{journal}{Phys. Rev. Lett.} \textbf{\bibinfo{volume}{8703}},
  \bibinfo{pages}{037004} (\bibinfo{year}{2001}).

\bibitem[{\citenamefont{Yip}(2002)}]{Yip2002}
\bibinfo{author}{\bibfnamefont{S.~K.} \bibnamefont{Yip}},
  \bibinfo{journal}{Phys. Rev. B} \textbf{\bibinfo{volume}{65}},
  \bibinfo{pages}{144508} (\bibinfo{year}{2002}).

\bibitem[{\citenamefont{Maki}(1966)}]{Maki1966}
\bibinfo{author}{\bibfnamefont{K.}~\bibnamefont{Maki}}, \bibinfo{journal}{Phys.
  Rev.} \textbf{\bibinfo{volume}{148}}, \bibinfo{pages}{362}
  (\bibinfo{year}{1966}).

\bibitem[{\citenamefont{Aoi et~al.}(1974)\citenamefont{Aoi, Meservey, and
  Tedrow}}]{AOI1974}
\bibinfo{author}{\bibfnamefont{K.}~\bibnamefont{Aoi}},
  \bibinfo{author}{\bibfnamefont{R.}~\bibnamefont{Meservey}}, \bibnamefont{and}
  \bibinfo{author}{\bibfnamefont{P.~M.} \bibnamefont{Tedrow}},
  \bibinfo{journal}{Phys. Rev. B} \textbf{\bibinfo{volume}{9}},
  \bibinfo{pages}{875} (\bibinfo{year}{1974}).

\bibitem[{\citenamefont{Clogston}(1962)}]{Clogston1962}
\bibinfo{author}{\bibfnamefont{A.~M.} \bibnamefont{Clogston}},
  \bibinfo{journal}{Phys. Rev. Lett.} \textbf{\bibinfo{volume}{9}},
  \bibinfo{pages}{266} (\bibinfo{year}{1962}).

\bibitem[{\citenamefont{Chandrasekhar}(1962)}]{Chandrasekhar1962}
\bibinfo{author}{\bibfnamefont{B.~S.} \bibnamefont{Chandrasekhar}},
  \bibinfo{journal}{Appl. Phys. Lett.} \textbf{\bibinfo{volume}{1}},
  \bibinfo{pages}{7} (\bibinfo{year}{1962}).

\bibitem[{SPs()}]{SPsolution}
\bibinfo{note}{The $b=0$ solution of the Maki equation actually represents the
  supercooling critical field of an intrinsically hysteretic first-order phase
  transition.}

\bibitem[{\citenamefont{Tedrow and Meservey}(1979)}]{Tedrow1979}
\bibinfo{author}{\bibfnamefont{P.~M.} \bibnamefont{Tedrow}} \bibnamefont{and}
  \bibinfo{author}{\bibfnamefont{R.}~\bibnamefont{Meservey}},
  \bibinfo{journal}{Phys. Rev. Lett.} \textbf{\bibinfo{volume}{43}},
  \bibinfo{pages}{384} (\bibinfo{year}{1979}).

\bibitem[{\citenamefont{Adams}(2004)}]{Adams2004}
\bibinfo{author}{\bibfnamefont{P.~W.} \bibnamefont{Adams}},
  \bibinfo{journal}{Phys. Rev. Lett.} \textbf{\bibinfo{volume}{92}},
  \bibinfo{pages}{067003} (\bibinfo{year}{2004}).

\bibitem[{\citenamefont{Adams et~al.}(1998)\citenamefont{Adams, Herron, and
  Meletis}}]{Adams1998}
\bibinfo{author}{\bibfnamefont{P.~W.} \bibnamefont{Adams}},
  \bibinfo{author}{\bibfnamefont{P.}~\bibnamefont{Herron}}, \bibnamefont{and}
  \bibinfo{author}{\bibfnamefont{E.~I.} \bibnamefont{Meletis}},
  \bibinfo{journal}{Phys. Rev. B} \textbf{\bibinfo{volume}{58}},
  \bibinfo{pages}{R2952} (\bibinfo{year}{1998}).

\bibitem[{\citenamefont{Tedrow and Meservey}(1982)}]{Tedrow1982}
\bibinfo{author}{\bibfnamefont{P.~M.} \bibnamefont{Tedrow}} \bibnamefont{and}
  \bibinfo{author}{\bibfnamefont{R.}~\bibnamefont{Meservey}},
  \bibinfo{journal}{Phys. Rev. B} \textbf{\bibinfo{volume}{25}},
  \bibinfo{pages}{171} (\bibinfo{year}{1982}).

\bibitem[{\citenamefont{Butko et~al.}(2000)\citenamefont{Butko, DiTusa, and
  Adams}}]{Adams2000b}
\bibinfo{author}{\bibfnamefont{V.~Y.} \bibnamefont{Butko}},
  \bibinfo{author}{\bibfnamefont{J.~F.} \bibnamefont{DiTusa}},
  \bibnamefont{and} \bibinfo{author}{\bibfnamefont{P.~W.} \bibnamefont{Adams}},
  \bibinfo{journal}{Phys. Rev. Lett.} \textbf{\bibinfo{volume}{84}},
  \bibinfo{pages}{1543} (\bibinfo{year}{2000}).

\bibitem[{\citenamefont{Bergmann and Horriaresser}(1985)}]{BERGMANN1985}
\bibinfo{author}{\bibfnamefont{G.}~\bibnamefont{Bergmann}} \bibnamefont{and}
  \bibinfo{author}{\bibfnamefont{C.}~\bibnamefont{Horriaresser}},
  \bibinfo{journal}{Phys. Rev. B} \textbf{\bibinfo{volume}{31}},
  \bibinfo{pages}{1161} (\bibinfo{year}{1985}).

\bibitem[{\citenamefont{Bergmann}(2001)}]{Bergmann2001}
\bibinfo{author}{\bibfnamefont{G.}~\bibnamefont{Bergmann}},
  \bibinfo{journal}{Phys. Rev. B} \textbf{\bibinfo{volume}{6319}},
  \bibinfo{pages}{193101} (\bibinfo{year}{2001}).

\end{thebibliography}

\newpage

\begin{figure}
\includegraphics[width=6in]{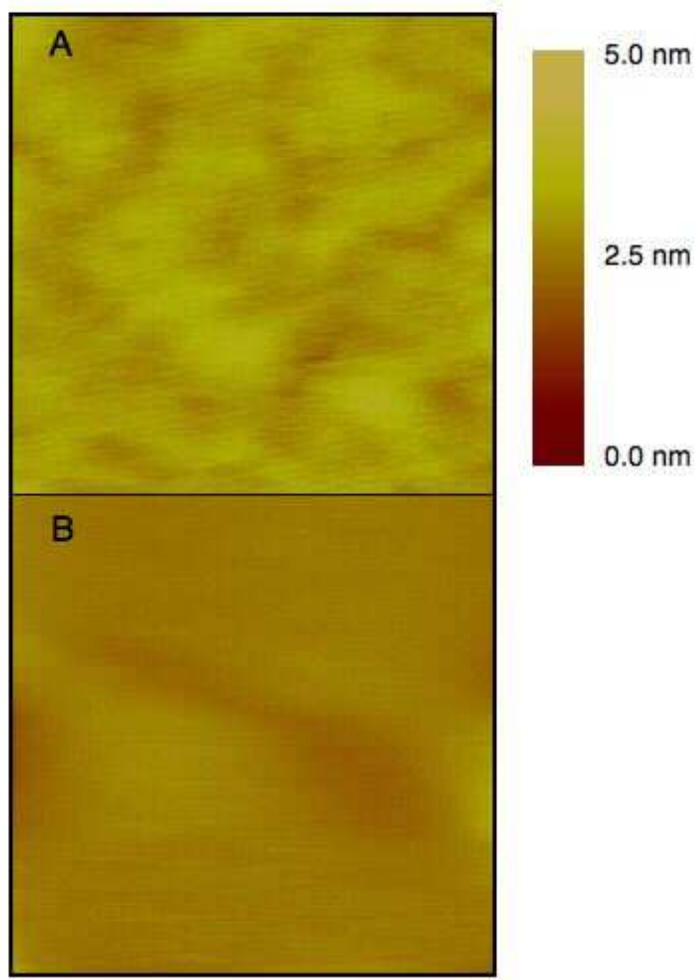}
\caption{\label{AFM} A) 0.1 x 0.1 $\mu$m atomic force micrograph of a 6 nm thick Be film evaporated onto glass at 84 K.  B) Micrograph of a Be film coated with 0.5 nm of Au. }
\end{figure}

\begin{figure}
\includegraphics[width=6in]{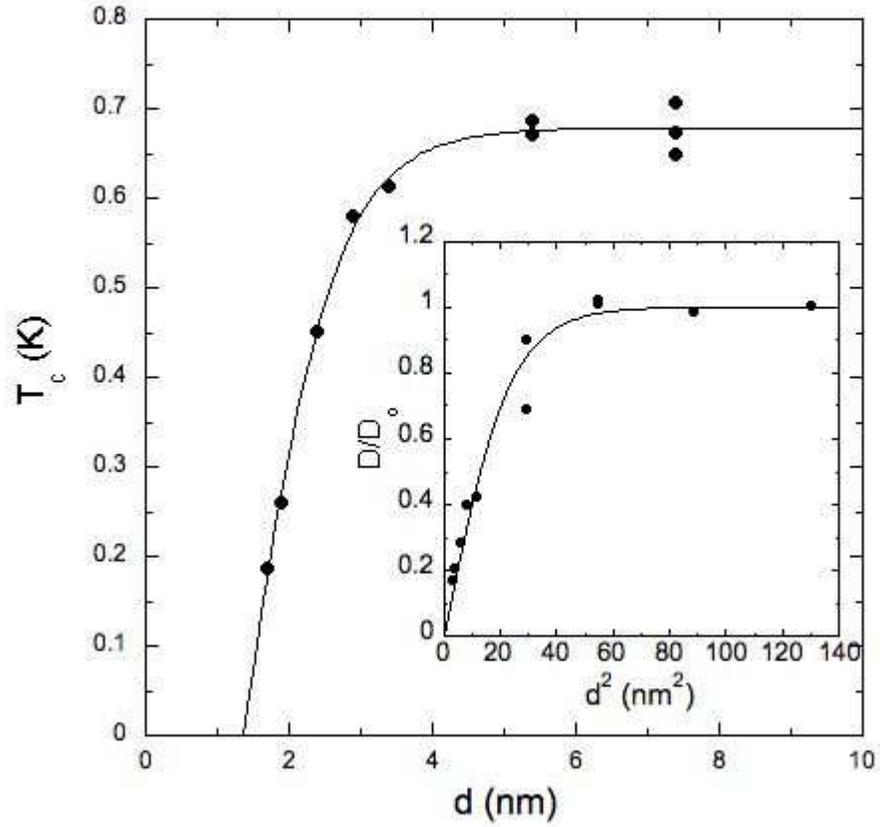}
\caption{\label{Tc&D} Transition temperature of Be/Au bilayers as a function of Be thickness.  The Au thickness was 0.5 nm in each sample.  The solid line is a best fit to the data using the empirical form of \req{empirical:Tc}.  Inset: relative diffusivity of Be/Au bilayers as a function of Be thickness.  The solid line is a fit to \req{empirical:D}. }
\end{figure}

\begin{figure}
\includegraphics[width=6in]{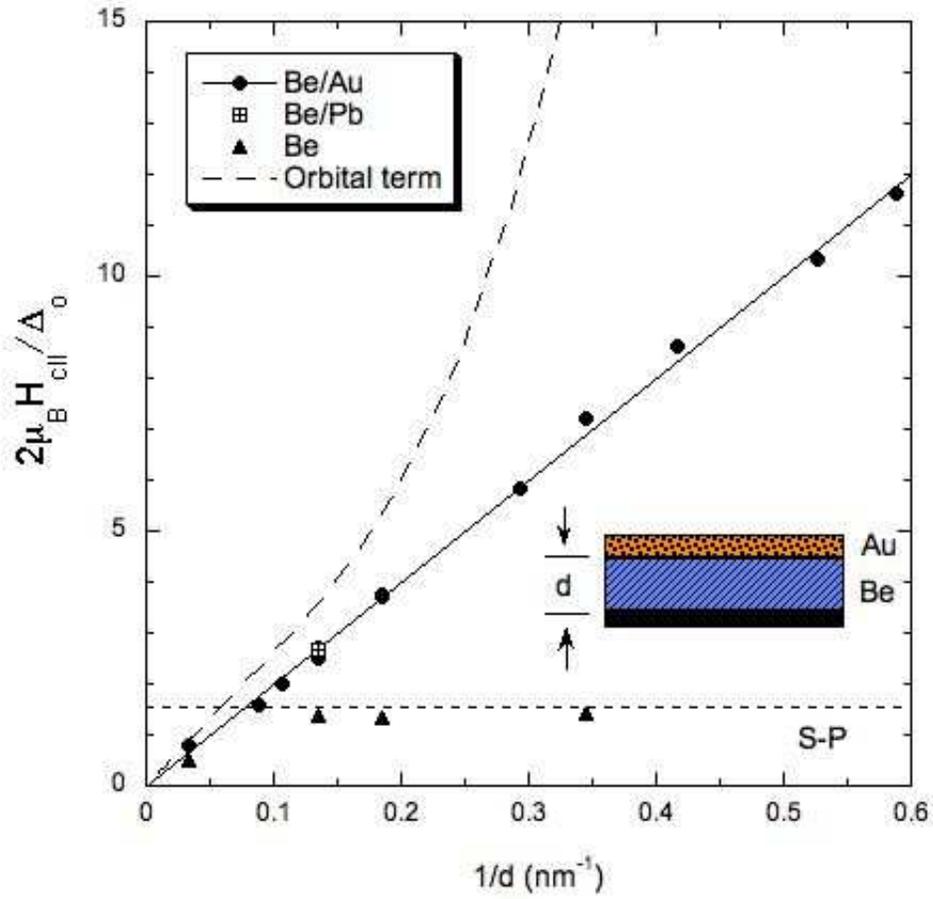}
\caption{\label{Linear} Normalized parallel critical fields as a function of Be thickness for Be/Au bilayers (circles), Be/Pb (crossed box), pure Be films (triangles).  The long dashed line represents the theoretical orbitally limited critical field given by \req{orbital}.  The solid line is a linear least fit to the bilayer data.  The horizontal dashed line represents the Clogston critical field. }
\end{figure}

\begin{figure}
\includegraphics[width=6in]{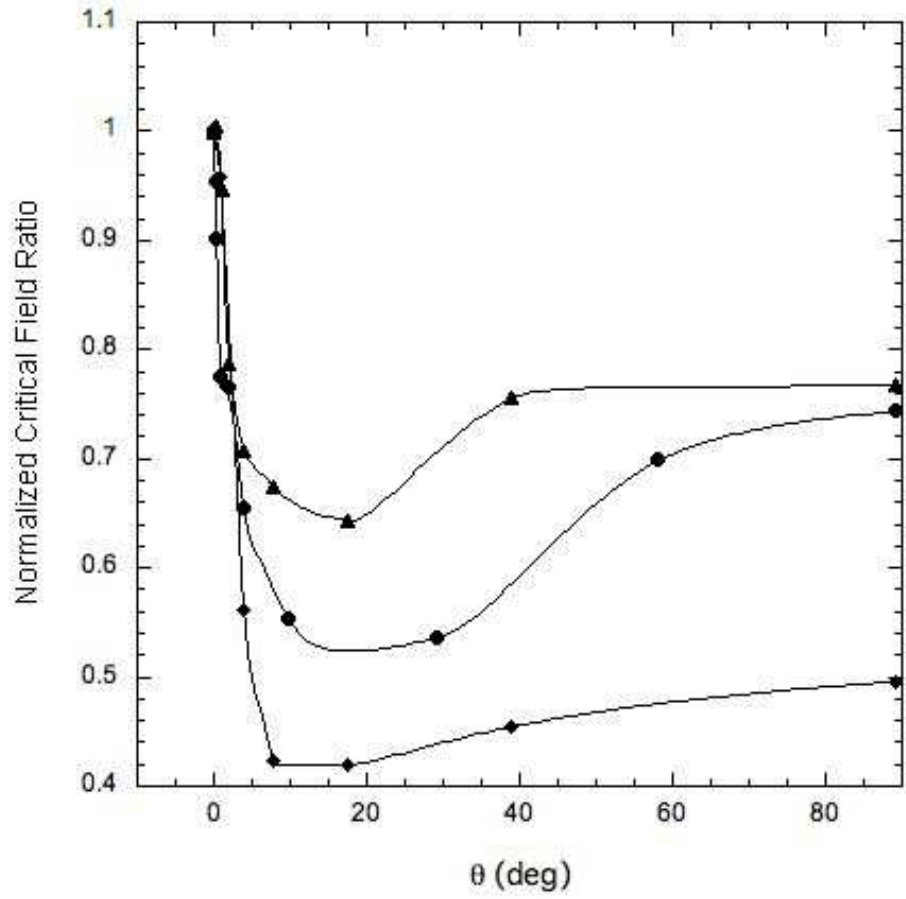}
\caption{\label{Dip} Ratio of the critical field of Be/Au bilayers and Be films of equal Be thickness as a function of tilt angle, $\theta=0$ corresponds to parallel field.  triangles: $d=7.4$ nm, circles: $d=5.4$ nm, diamonds: $d=2.9$ nm.  The solid lines are a guide to the eye.}
\end{figure}

\begin{figure}
\includegraphics[width=6in]{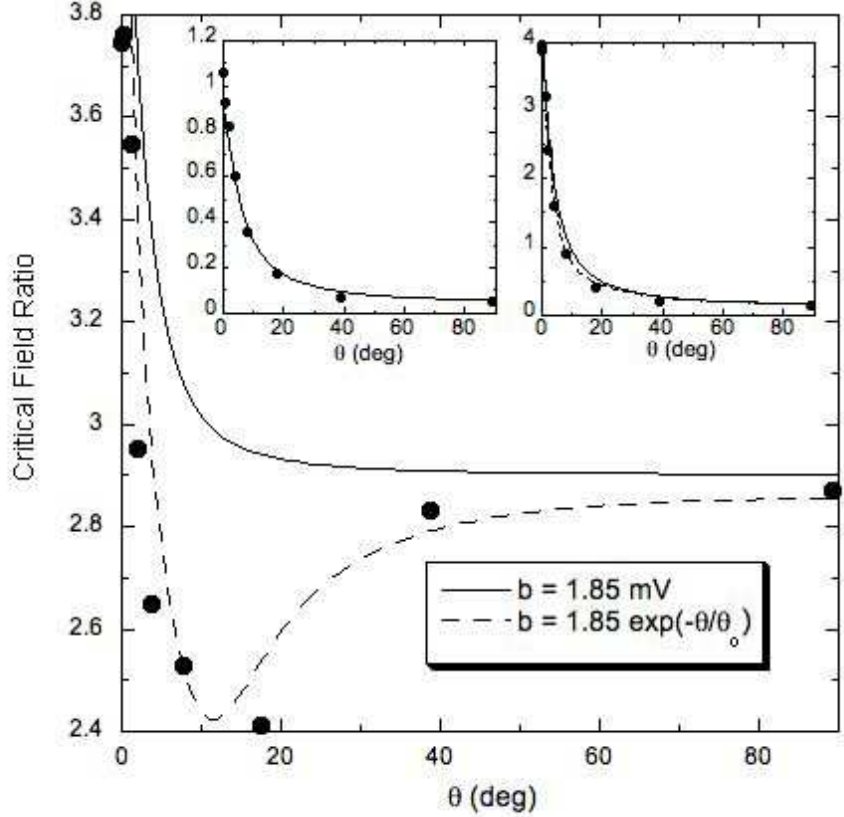}
\caption{\label{DipFit} Ratio of the critical field of a 5.4 nm Be/Au bilayer ($T_c=0.68$ K, $H_{c2}=0.137$ T, $R=240\;\Omega$) and 5.4 nm Be film ($T_c=0.505$ K, $H_{c2}=0.048$ T, $R=162\;\Omega$) as a function of tilt angle.  Solid line is the solution of \req{Maki} assuming an isotropic SO coupling parameter of $b=1.85$ mV for the Be/Au bilayer and $b=0.0132$ mV for the Be film.  The dashed line is the solution of \req{Maki} assuming that the Be/Au SO parameter is exponentially attenuated with increasing tilt angle.  Left inset:  Fit of \req{Maki} to the Be film data.  Right inset: Solid line is a fit of \req{Maki} to the Be/Au data with a constant $b$.  The dashed line is a somewhat better fit using an exponentially form $b=b_o\exp(-\theta/\theta_o)$.}
\end{figure}



\end{document}